\documentclass[aps,pre,twocolumn,superscriptaddress,showpacs,showkeys,amsmath,amssymb,floatfix]{revtex4-2}
\usepackage{epsfig,amsmath,amssymb,xcolor,bm}
\usepackage[T1]{fontenc}

\begin{document}

  \title{Electric-field driven flat bands in the distorted sawtooth chain via the Katsura-Nagaosa-Balatsky mechanism}

  \author{Vadim Ohanyan}
  \affiliation{Laboratory of Theoretical Physics
    Yerevan State University,
    1 Alex Manoogian Str., 0025 Yerevan, Armenia}
  \affiliation{CANDLE, Synchrotron Research Institute, 31 Acharyan Str., 0040 Yerevan, Armenia}

  \author{Lusik Amiraghyan}
  \affiliation{Laboratory of Theoretical Physics
    Yerevan State University,
    1 Alex Manoogian Str., 0025 Yerevan, Armenia}
  \affiliation{Institute of Applied Problems of Physics, 25 Hr. Nersisyan St, Yerevan 0014, Armenia}

  \author{Michael Sekania}
  \affiliation{Rechenzentrum, University of Augsburg, 86135 Augsburg, Germany }
  \affiliation{Andronikashvili Institute of Physics, Javakhishvili Tbilisi State University, Tamarashvili str. 6, 0177 Tbilisi, Georgia}

  \author{Marcus Kollar}
  \affiliation{Theoretical Physics III, Center for Electronic Correlations and Magnetism, Institute of Physics,
    University of Augsburg, 86135 Augsburg, Germany}

  \date{\today}

  \begin{abstract}
    We investigate flat magnonic bands in a generalized sawtooth-chain
    model in which three sets of exchange parameters (symmetric
    Heisenberg exchange, axial Ising anisotropy, and antisymmetric
    Dzyaloshinskii-Moriya (DM) exchange) are assigned independently to
    each side of the triangular plaquette. If the effective
    Dzyaloshinskii-Moriya (DM) interaction parameters are generated
    via the Katsura-Nagaosa-Balatsky (KNB) mechanism of
    magnetoelectricity, they become explicit functions of the
    electric-field magnitude and direction, as well as of the lattice
    geometry, which in the present casen is characterized by two bond
    angles. We focus on the situation in which these two angles are
    unequal, corresponding to a distortion of the triangular
    plaquette. Several electric-field induced flat-band scenarios in
    the distorted sawtooth chain are analyzed, and expressions are
    derived for the electric-field strength required to drive the
    one-magnon excitations into a flat-band regime when the field is
    aligned along the lattice bonds. The saturation field and its
    dependence on the distortion angle are also examined. Finally, we
    establish a mapping between the flat-band solutions for a general
    DM interaction and its specific KNB-induced form.
    \\~\\
    \emph{This article is dedicated to the memory of Johannes Richter.}
  \end{abstract}

  \pacs{71.10.-w,
    75.10.Lp,
    75.10.Jm}

  \keywords{Sawtooth chain, KNB mechanism, Magnetoelectric effect, Localized magnons,  Flat bands}

  \maketitle

  \section{\label{sec:1}Introduction}

  Localization in many-body quantum models continues to be a focus of
  numerous researches.  The most extensively studied many-body quantum
  models exhibiting well-understood localized states are tight-binding
  models, or quantum lattice models with mobile non-interacting
  particles \cite{der15, ley18, bae23, lee24, mall25}. The first
  evidence for dispersionless single-particle eigenstates in a
  two-dimensional tight-binding Hamiltonian appeared in the 1980s
  \cite{suth86}. Shortly thereafter, the analysis of special lattice
  topologies within the Hubbard model framework led to the prediction
  of interaction-driven ferromagnetism at half-filling, even for
  finite on-site repulsion $U$ \cite{mielke, tas92, mielke93, tas98,
    Moes10, Moes12, baliha}. More recent developments have described
  the paramagnetic-to-ferromagnetic transition in flat-band Hubbard
  systems in terms of Pauli-correlated percolation \cite{Moes10,
    Moes12, baliha}. In the past decades, significant experimental
  progress was made in realizing flat-band physics in photonic
  lattices \cite{Ley18a, nat24, phot25}, ultracold atoms in optical
  lattices \cite{Taie15, Ozawa17, Taie20}, twisted bilayer graphene
  \cite{gr23, gr24}, and other topological flat-band materials
  \cite{tfb1,tfb2}. By contrast, Heisenberg and other quantum spin
  Hamiltonians are intrinsically strongly interacting systems, and
  localized states in such translationally invariant spin models might
  initially appear to belong to the class of many-body localized (MBL)
  states. Indeed, MBL physics is well-established in disordered
  one-dimensional quantum spin models \cite{MBL}. However, the
  phenomenon of localized magnons in frustrated magnets shares many
  conceptual similarities with flat-band localization in tight-binding
  systems \cite{schul02, rich04, rich05, zhi04, der04, sc_KNB,
    JJJ}. Among the various excitations in quantum spin models the
  magnonic excitations are special, in that they occur as low-lying
  quasi-particle excitation above a saturated state in the spectrum of
  all such models.
  \begin{figure}[tb]
    \begin{center}
      \includegraphics[width=75.5mm]{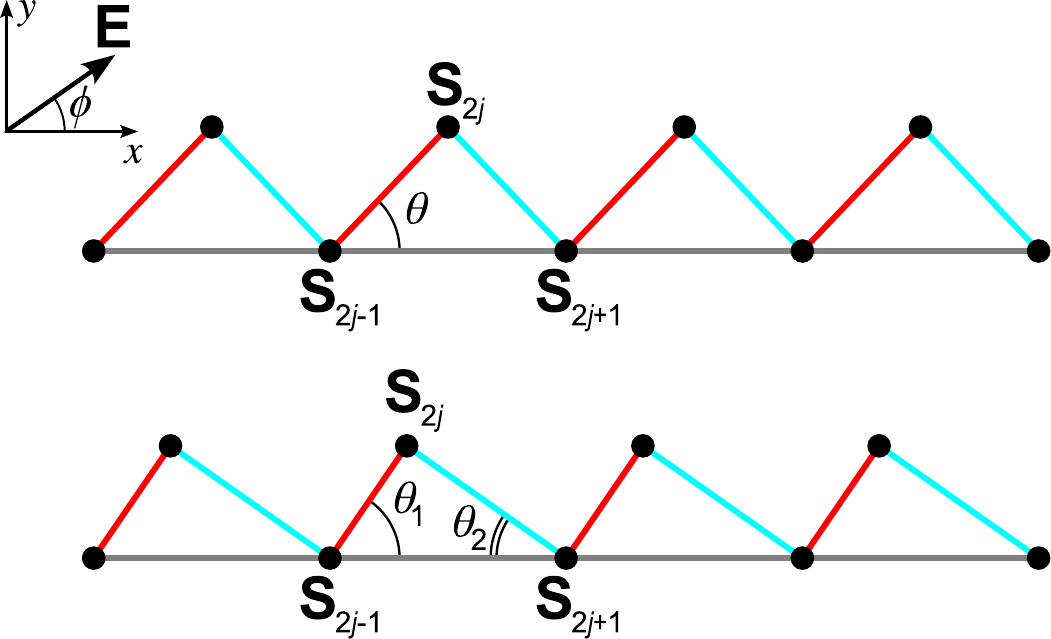}
      \caption{(Color online) Symmetric and distorted sawtooth chains
        with three different coupling(s) for each side of
        triangle. The filled circles show the lattice sites occupied
        by spins. The exchange coupling, the $XXZ$ anisotropy and DM
        interaction along the basal line (black) are $J_1$, $\Delta_1$
        and $D_1$, respectively. Left (red) and right (blue) bonds
        along the zigzag line feature the parameters $J_2$, $\Delta_2$
        and $D_2$ (red) and $J_3$, $\Delta_3$ and $D_3$ (blue),
        respectively. The symmetric case features the same bond angles
        $\theta$ for the left and right bonds of the triangle, while
        in the distorted chain they are $\theta_1$ and $\theta_2$,
        respectively. The basal line is chosen to be parallel to
        $x$-axis and electric field vector lies in $(x, y)$ plane. }
      \label{fig1}
    \end{center}
  \end{figure}
  When the magnetic field magnitude is close to and below the
  saturation field the main magneto-thermal properties of the system
  can be captured by the model of non-interacting magnons, which has
  the form of a tight-binding model. The corresponding Hamiltonian can
  be constructed within linear spin-wave theory in terms of the
  Holstein-Primakoff transformation \cite{JJJ}. Alternatively, for
  each quantum-spin lattice model it is straightforward to construct a
  one-magnon state, as a translationally invariant superposition of
  $S_j^-\left|0\right\rangle$, where
  $\left|0\right\rangle=\left|S, S,..., S\right\rangle$ is the
  reference polarized or ferromagnetic state, a direct product of
  $S^z=S$ spin states at each sites of the lattice. The eighenvalues
  of the Hamiltonian for such a states are the branches of the exact
  one-magnon spectrum, with the number of branches corresponding to
  the number of the site in the unit cell. A number of lattice spin
  models demonstrating localized magnons are known in all dimensions
  \cite{der15}. However, almost all of them demand additional
  constraints for the exchange constants of the model, transforming
  the lower band of the one-magnon spectrum into a flat one. One of
  the simplest quantum spin chains which admit localized magnons is the
  sawtooth chain \cite{der15, schul02, rich04, rich05, zhi04, der04,
    sc_KNB, JJJ}, which is   a chain of corner-sharing
  triangles (see Fig.~\ref{fig1}). The $XXZ$ model on the sawtooth chain
  with two exchange couplings, $J_1$ for the basal line and $J_2$ for
  the zigzag part, exhibits a flat band if
  \begin{eqnarray}\label{eq:FB_J1J2}
    J_2=\pm\sqrt{2(1+\Delta_1)}J_1,
  \end{eqnarray}
  where $\Delta_1$ is the axial exchange anisotropy for the basal line
  \cite{schul02}. However, this restriction can be overcome with the
  aid of electric field \cite{sc_KNB, JJJ} in the sawtooth chain by a
  special type  of magnetoelectric coupling, the
  Katsura-Nagaosa-Balatsky (KNB) mechanism \cite{KNB1, KNB2, Sol21,
    Sol25}. It was shown in Ref.~\cite{sc_KNB} that in the sawtooth
  chain with two exchange couplings and no axial anisotropy
  ($\Delta_1=1$), the lower branch of the one-magnon spectrum becomes
  flat when electric field is pointed along the basal line and its
  magnitude is given by
  \begin{eqnarray}\label{eq:E_FB}
  E_{FB}=\pm\frac{\sqrt{4J_1^2-J_2^2}}{\sin\theta}.
  \end{eqnarray}
  This result is quite remarkable, as it applies to arbitrary values
  of the exchange constants, provided $4J_1^2>J_2^2$. Here $\theta$ is
  the angle between the basal line and zigzag bonds in case of
  isosceles triangles. The KNB mechanism and an external in-plane
  electric field induce additional Dzyaloshinskii-Moriya (DM)
  interactions, which in general are different for the three sides of
  the triangle. The values of this induced DM exchange depend on the
  electric field magnitude $E$, its angle with the basal line $\phi$
  and the lattice bond angle $\theta$. Further developments of the
  ideas and results of Ref.~\cite{sc_KNB} were presented in
  Ref.~\cite{JJJ}, where the generalized model of an $XXZ$ sawtooth
  chain with three symmetric exchange couplings and three DM
  interactions was considered. The flat-band constraints derived in
  Ref.~\cite{JJJ} are a system of two rational equations supplemented
  with an additional conditions of band ordering. The number of
  parameters involved in the constraints is seven, namely three
  symmetric exchange coupling, three DM parameters, and the axial
  anisotropy on the basal line. Therefore, a number of possible
  solution schemes were considered in Ref.~\cite{JJJ}. Each scheme
  offers a solution in terms of two or three quantities out of seven,
  when the remaining six or five are assumed to be given. Particularly
  interesting was a case for which the three DM parameters were given
  in terms of three symmetric exchange couplings and the axial
  anisotropy on the basal line.  Parametrized by one continuous
  parameter, an infinite number of flat-band solutions was
  obtained. Moreover, it was also shown in Ref.~\cite{JJJ} how to
  obtain flat-band solutions in terms of DM parameters induced by the
  KNB mechanism, i.e., by tuning the angle and the magnitude of the
  electric field. Nowadays, research on various aspects of spin
  ordering and one-magnon dynamics in the sawtooth chain with
  inhomogeneous couplings remains very active \cite{bren20, hutak,
    noner, rausch23, reich, rausch25}.

  In the present paper we continue our investigations into the nature
  and features of the electric-field driven flat bansd in the sawtooth
  chain. We generalize the model, considering now the sawtooth chain
  with distorted triangles. Namely the angles at the base line of the
  triangle, $\theta_1$ and $\theta_2$, are different (see
  Fig.~\ref{fig1}), which arises naturally from unequal length of
  these bonds.  Our main finding is that, compared to the
  bond-symmetric case \cite{sc_KNB}, a flat band can be induced with
  more flexibility in the case of $\theta_2 \neq \theta_1$, as
  detailed below.

  \section{\label{sec:3}The model and KNB polarization}

  As in our previous paper \cite{JJJ} here we deal with the model of
  an $XXZ$ sawtooth chain with DM interaction and three sets of
  interaction parameters, one for each side of triangle. The
  Hamiltonian thus has the following form,
  \begin{widetext}
    \begin{eqnarray}\label{eq:ham_gen}
      \mathcal{H}&=&\sum_{j=1}^{N/2}\left(\frac{J_1+iD_1}{2}S_{2j-1}^{+}S_{2j+1}^{-}+\frac{J_1-iD_1}{2}S_{2j-1}^{-}S_{2j+1}^{+}+J_1\Delta_1 S_{2j-1}^{z}S_{2j+1}^{z}\right)\\
                 &+&\sum_{j=1}^{N/2}\left(\frac{J_2+iD_2}{2}S_{2j-1}^{+}S_{2j}^{-}+\frac{J_2-iD_2}{2}S_{2j-1}^{-}S_{2j}^{+}+J_2\Delta_2 S_{2j-1}^{z}S_{2j}^{z}\right)\nonumber\\
                 &+&\sum_{j=1}^{N/2}\left(\frac{J_3+iD_3}{2}S_{2j}^{+}S_{2j+1}^{-}+\frac{J_3-iD_3}{2}S_{2j}^{-}S_{2j+1}^{+}+J_3\Delta_3 S_{2j}^{z}S_{2j+1}^{z}\right)-B\sum_{j=1}^N S_j^z, \nonumber
    \end{eqnarray}
  \end{widetext}
  where $S_j^{\alpha}$, $\alpha=x, y, z$ and
  $S_j{\pm}=S_j^x \pm i S_j^y$ stand for the standard $SU(2)$
  generators with
  $[S_j^{\alpha},\,S_{j^{\prime}}^{\beta}] =
  \delta_{jj^{\prime}}\epsilon^{\alpha\beta\gamma}S_{j}^{\gamma}$. In
  order to have conservation of $S^z_{tot}=\sum_{j=1}^N S^z_j$, the
  direction of the Dzyaloshinskii vector coincides with the direction
  of the magnetic field. As mentioned above, the spin-induced
  ferroelectricity scheme of the KNB mechanism leads to DM terms, as
  the microscopic polarization of the bond between two spin
  \cite{KNB1} can be expressed as
  \begin{eqnarray}\label{eq:KNB_P}
    \bm{P}_{j,j+1}=\gamma_{j,j+1}\bm{e}_{j,j+1}\times\left(\bm{S}_j\times\bm{S}_{j+1}\right),
  \end{eqnarray}
  where $\bm{e}_{j,j+1}$ is the unit vector pointed form site $j$ to
  the site $j+1$ and $\gamma_{j,j+1}$ is a constant depending on the
  local features of the chemical bonds. The polarization appears due
  to a complicated redistribution of the electron density between two
  magnetic ions and the $p$-element atom in between which is
  covalently bonded to them\cite{KNB1,KNB2,Sol21, Sol25}. The KNB
  polarization of the model includes an additional contribution to the
  Hamiltonian when an external electric field is applied,
  \begin{eqnarray}\label{eq:KNB_H}
    \mathcal{H}_{P}=-\bm{E}\cdot\sum_{\left(i, j\right)}\gamma_{i,j}\,\bm{e}_{i,j}\times\left(\bm{S}_i\times\bm{S}_{j}\right).
  \end{eqnarray}
  Clearly this Hamiltonian can be represented in a general DM form,
  \begin{eqnarray}
    \mathcal{H}_{P}&=&\sum_{\left(i, j\right)}\bm{D}_{i,j}(\bm{E})\cdot\left(\bm{S}_i\times\bm{S}_{j}\right), \\
    \bm{D}_{i,j}(\bm{E})&=&-\gamma_{ij}\left(\bm{E}\times\bm{e}_{ij}\right). \nonumber
  \end{eqnarray}
  Here the electric-field dependent DM vector
  $\bm{D}_{i,j}\left(\bm{E}\right)$ can be chosen parallel to the
  $z$-axis for the entire lattice, provided it is planar, i.e., for
  all bonds
  \begin{eqnarray}\label{eq:e_ij}
    \bm{e}_{ij}=A_{ij}\bm{e}_x+B_{ij}\bm{e}_y
  \end{eqnarray}
  and the electric field vector then lies in the same plane. Thus in
  order to obtain a quantum spin model with conserved $S_{tot}^z$,
  only special 1d and 2d lattices can be considered. The simplest case
  is a linear chain leading to \cite{bro13,oles, mench15, thakur18,
    sznajd18, XYZ, sznajd19,oha20}
  \begin{eqnarray}\label{eq:KNB_lin}
    P^x&=&0 \\
    P^y&=&\gamma\sum_{j=1}^N\left(S_J^yS_{j+1}^x-S_{j}^x S_{j+1}^y\right) \nonumber \\
    P^z&=&\gamma\sum_{j=1}^N\left(S_J^zS_{j+1}^x-S_{j}^x S_{j+1}^z\right) \nonumber.
  \end{eqnarray}
  The next example is the spin chain folded to a regular zigzag
  structure, for which
  $\bm{e}_{j}=\cos\theta\bm{e}_x+\left(-1\right)^{j-1}
  \sin\theta\bm{e}_y$ and thus \cite{baran18,jap19,jap21, sc_KNB, JJJ,
    bar21}
  \begin{eqnarray}
    &-&\bm{E}\cdot\bm{P}\\
    &=&
        \sum_{j=1}^N\left(E_y\cos\theta+({-}1)^j E_x\sin\theta\right)\left(S_{j}^xS_{j+1}^y {-} S_{j}^yS_{j+1}^x\right).\nonumber
  \end{eqnarray}
  Here the constant $\gamma$ is absorbed into the definition of
  renormalized electric field components, $E_x$ and $E_y$, measured in
  appropriate units. The symmetric sawtooth chain
  ($\theta_2=\theta_1$) contains both a linear chain and a zigzag
  part, thus in this case the interaction between in-plane electric
  field and KNB polarization has the following form \cite{sc_KNB}.,
  \begin{eqnarray}\label{eq:KNB_sawtooth}
    &-&\bm{E}\cdot\bm{P}\\
    &=&\sum_{j=1}^N\left( E_y\cos\theta+(-1)^j E_x\sin\theta\right)\left(S_j^xS_{j+1}^y-S_j^yS_{j+1}^x\right)\nonumber \\
    &+&a E_ y
        \sum_{j=1}^{N/2}\left(S_{2j-1}^xS_{2j+1}^y-S_{2j-1}^yS_{2j+1}^x\right). \nonumber
  \end{eqnarray}
  The coefficient $a$ here accounts for the possible difference in
  the microscopic features of the KNB mechanism for the bonds in the basal
  line and the zigzag part of the chain.  It is easy to see  that in
  case of an  in-plane electric field, the interaction induced by the  KNB
  polarization  contributes a DM-type interaction as in Eq.~(\ref{eq:ham_gen}) with
  \begin{eqnarray}\label{eq:DDD}
    D_1&=&aE\sin\phi,\\
    D_2&=&E\sin\left(\phi-\theta\right), \nonumber \\
    D_3&=&E\sin\left(\phi+\theta\right), \nonumber
  \end{eqnarray}
  where $\phi$ is the angle between the electric field and the basal line ($x$
  axis) of the chain. This case was analyzed in detail in
  Ref.~\cite{sc_KNB} and \cite{JJJ}. Finally for the distorted
  sawtooth chain (see Fig.~\ref{fig1}), which we analyze in the
  following, the KNB mechanism for an in-plane electric field gives a
  Hamiltonian of the form (\ref{eq:ham_gen}) with DM parameters given
  by
  \begin{eqnarray}\label{eq:DDD_def}
    D_1&=&aE\sin\phi,\\
    D_2&=&E\sin\left(\phi-\theta_1\right), \nonumber \\
    D_3&=&E\sin\left(\phi+\theta_2\right), \nonumber
  \end{eqnarray}
  The general flat-band constraints derived and analyzed in
  Ref.~\cite{JJJ} still apply. However now the connetion between DM
  parameters and electric field is more complicated and contains an
  additional parameter, the deformation angle of the sawtooth
  chain. We note that models of the KNB mechanism in $2d$ and $3d$
  magnets have also been in focus recently \cite{esa, bre1, bre2}.

  \section{Flat-band constraints and general solutions}
  As the flat-band constraints and the properties of their solution do
  not depend on the magnitude of spin, we therefore without loss of
  generality consider here only the case $S=1/2$. The one-magnon
  spectrum and the corresponding flat-band constraints obtained in
  \cite{JJJ} are
  \begin{eqnarray}\label{1m_sp_gen}
    \varepsilon^{\pm}_1(k)&=&B-\frac 12\sum_{a=1}^3J_a\Delta_a+\frac 12\rho_1\cos(k-\phi_1)\\
                          &&\pm\frac12\big[\left(J_1\Delta_1-\rho_1\cos(k-\phi_1)\right)^2\nonumber\\
                          &&~~~~~~~+2\rho_2\rho_3\cos(k-\phi_2-\phi_3)+\rho_2^2+\rho_3^2\big]^{\frac12}\,,\nonumber
  \end{eqnarray}
  and
  \begin{eqnarray}\label{FB_pr}
    \phi_1&=&\phi_2+\phi_3\pm \pi \delta_{\sigma, -1}, \\
    0&=&\rho_1^2\left(\rho_2^2+\rho_3^2\right)+ 2\sigma J_1\Delta_1\rho_1\rho_2\rho_3-\rho_2^2\rho_3^2\,,\nonumber \\
    \sigma&=&\mbox{sign}\left(1+\frac{D_1D_2}{J_1J_2}\right)\equiv \mbox{sign}\left(1+\frac{D_1D_3}{J_1J_3}\right) \nonumber \\
    &\equiv& \mbox{sign}\left(1-\frac{D_2D_3}{J_2J_3}\right)=1 \nonumber
  \end{eqnarray}
  respectively, where the following notations were adopted,
  \begin{eqnarray}\label{eq:notations}
    J_a+i D_a &=&\rho_a e^{i \phi_a},\; \rho_a=\sqrt{J_a^2+D_a^2}, \\
    \phi_a&=&\arctan\frac{D_a}{J_a}, \nonumber
  \end{eqnarray}
  After the substitution of the solution of flat-band constraints given in Eq.~(\ref{FB_pr}), the one-magnon spectrum (\ref{1m_sp_gen}) takes the following form:
  \begin{eqnarray}\label{eq:one-m_spec}
    \varepsilon^{+}_1(k)&=&B-E_0+\rho_1\cos\left(k-\phi_1\right), \\
    \varepsilon^{-}_1(k)&=&B-B_0, \nonumber
  \end{eqnarray}
  where
  \begin{eqnarray}\label{eq:EB0}
    E_0&=&\frac{1}{2} \left(2J_1\Delta_1+J_2\Delta_2+J_3\Delta_3-\frac{ \rho_2 \rho_3}{\rho_1}\right), \nonumber \\
    B_0&=&\frac{1}{2} \left(J_2\Delta_2+J_3\Delta_3+\frac{ \rho_2 \rho_3}{\rho_1}\right),
  \end{eqnarray}
  and $B_0$ is the magnetic saturation field.  The last condition in
  Eqs.~(\ref{FB_pr}) guarantees that the flat band lies below the
  dispersive one. Formally the flat-band constraints are a system of
  two rational equations for seven parameters,
  $J_1, J_2, J_3, D_1, D_2, D_3$ and $\Delta_1$.  Hence there is a
  broad variety of possible solutions, several classes of which were
  analyzed in Ref.~\cite{JJJ}. Solutions which are given in terms of the
  three DM constants are particularly interesting as they make a
  direct link with the electric-field driven flat band discussed in
  Ref.~\cite{sc_KNB}.  In that case one can express all three DM
  couplings in terms of the other Hamiltonian parameters in the
  following way,
  \begin{eqnarray}\label{sol_y=ax}
    D_1&=&\mp J_1\frac{2J^2\alpha\left(1+\alpha\right)\sqrt{\frac{1}{\alpha}+\frac{ W_{\mu}(\alpha)}{2J^2 \alpha^2}}}{ W_{\mu}(\alpha)},\\
    D_2&=&\pm J_2\sqrt{\frac{1}{\alpha}+\frac{ W_{\mu}(\alpha)}{2J^2 \alpha^2}}, \nonumber \\
    D_3&=&\pm J_3\alpha\sqrt{\frac{1}{\alpha}+\frac{ W_{\mu}(\alpha)}{2J^2\alpha^2}}, \nonumber
  \end{eqnarray}
  where
  \begin{eqnarray}\label{sol_y=ax_2}
    J&=&\frac{J_2J_3}{J_1}, \\
    W_{\mu}(\alpha)&=&R(\alpha)+\mu\sqrt{R^2(\alpha)+4 J^2 \alpha (1+\alpha)\left(J_2^2+\alpha J_3^2\right)}, \nonumber \\
    R(\alpha)&=&J_2^2+\alpha^2 J_3^2-2\alpha \xi \Delta_1 J_2 J_3, \nonumber \\
    \mu&=&\left\{-1,1\right\}. \nonumber
  \end{eqnarray}
  Taking into account the square roots and the condition for ordering
  of the bands we obtain an additional constraint for the function
  $W_{\mu}(\alpha)$,
  \begin{eqnarray}\label{W_const}
    \alpha&>&0, \;\;\; W_{\mu}(\alpha)\in \left(-2J^2\alpha, \, 0\right), \\
    \alpha&<&0, \;\;\; W_{\mu}(\alpha)\in \left(0, \, -2J^2\alpha\right).  \nonumber
  \end{eqnarray}
  As $W_{1}(\alpha)$ is more likely to be positive and
  $W_{-1}(\alpha)$ negative, it is easier to find flat bands with
  $\text{sign}(\mu\alpha)=-1$, although the other case is still
  possible. To simplify the situation we consider the important
  particular case of homogeneous symmetric exchange parameters,
  $J_1=J_2=J_3=J=1$ and $\Delta_1=\pm 1$, for which the DM parameters
  given in Eq.~(\ref{sol_y=ax_2}) for become
  \begin{eqnarray}
    D_1&=&\mp \frac{2\alpha\left(1+\alpha\right)\sqrt{\frac{1}{\alpha}+\frac{ X_{\mu}(\alpha)}{2 \alpha^2}}}{ X_{\mu}(\alpha)},\\
    D_2&=& \pm \sqrt{\frac{1}{\alpha}+\frac{ X_{\mu}(\alpha)}{2 \alpha^2}}, \nonumber \\
    D_3&=&\pm \alpha\sqrt{\frac{1}{\alpha}+\frac{ X_{\mu}(\alpha)}{2\alpha^2}}, \nonumber
  \end{eqnarray}
  where
  \begin{eqnarray}\label{Xmu}
    && X_{\mu}(\alpha)=\\
    &&\left\{
       \begin{array}{ll} (1-\alpha)^2+\mu\sqrt{\left(1-\alpha\right)^4+4\alpha \left(1+\alpha\right)^2}, \; \Delta_1=1\,, \\
         \left(1+\alpha\right)^2+\mu\sqrt{\left(1+\alpha\right)^4+4\alpha \left(1+\alpha\right)^2}, \; \Delta_1=-1\,.       \end{array}\right.  \nonumber
  \end{eqnarray}
  Here an additional constraint restricting the possible values of
  $\alpha$ (for given values of $\mu$) must be imposed here ensure a
  positive value of the radicand. For $\Delta_1=1$ only $\mu=1$ and
  $\alpha<0$ are acceptable, in which case the radicand in
  Eq.~(\ref{Xmu}) is always positive. For the case $\Delta_1=-1$ both
  signs of $\mu$ can be compatible with the flat bands condition,
  however, $\alpha$ still must be negative, and there is an additional
  forbidden range
  \begin{eqnarray}\label{al_range}
    \alpha \in \left(-3-2\sqrt 2, -3+2\sqrt 2\right).
  \end{eqnarray}
  for which the radicand is negative.

  It is possible to map general DM parameters to the KNB case, when
  they are given by the Eq.~(\ref{eq:DDD}).  As discussed in
  Ref.~\cite{JJJ}, if $D_1, D_2$ and $D_3$ are the solution of the
  flat-band constraint given in the Eqs.~(\ref{sol_y=ax})-(\ref{Xmu}),
  they can be induced by the KNB mechanism with the following
  magnitude and angle of the electric field and bond angle $\theta$,
  \begin{eqnarray}\label{eq:phi_E_2}
    E&=&\pm\frac{2|D_1|}{|a|}\sqrt{\frac{D_1^2-a^2D_2D_3}{4D_1^2-a^2(D_2+D_3)^2}}, \\
    \phi&=&\arctan\left(\frac{\text{sign}(D_3+D_2)}{|a|(D_3-D_2)}\sqrt{4D_1^2-a^2(D_2+D_3)^2}\right).\nonumber \\
    \theta&=&\arctan\frac{\sqrt{4D_1^2-a^2\left(D_2+D_3\right)^2}}{\left|a\left(D_2+D_3\right)\right|}.\nonumber
  \end{eqnarray}
  Here the ratio $a$ of the KNB constants for the basal line and
  zigzag bond of the lattice is arbitrary, and the only additional
  condition for the mapping is \cite{JJJ}
  \begin{eqnarray}\label{eq:4D1^2}
    4D_1^2-a^2(D_2+D_3)^2>0.
  \end{eqnarray}

  \section{Electric-field induced flat band in the distorted sawtooth chain}

\begin{figure}[b ]
    \begin{center}
      \includegraphics[width=75.5mm]{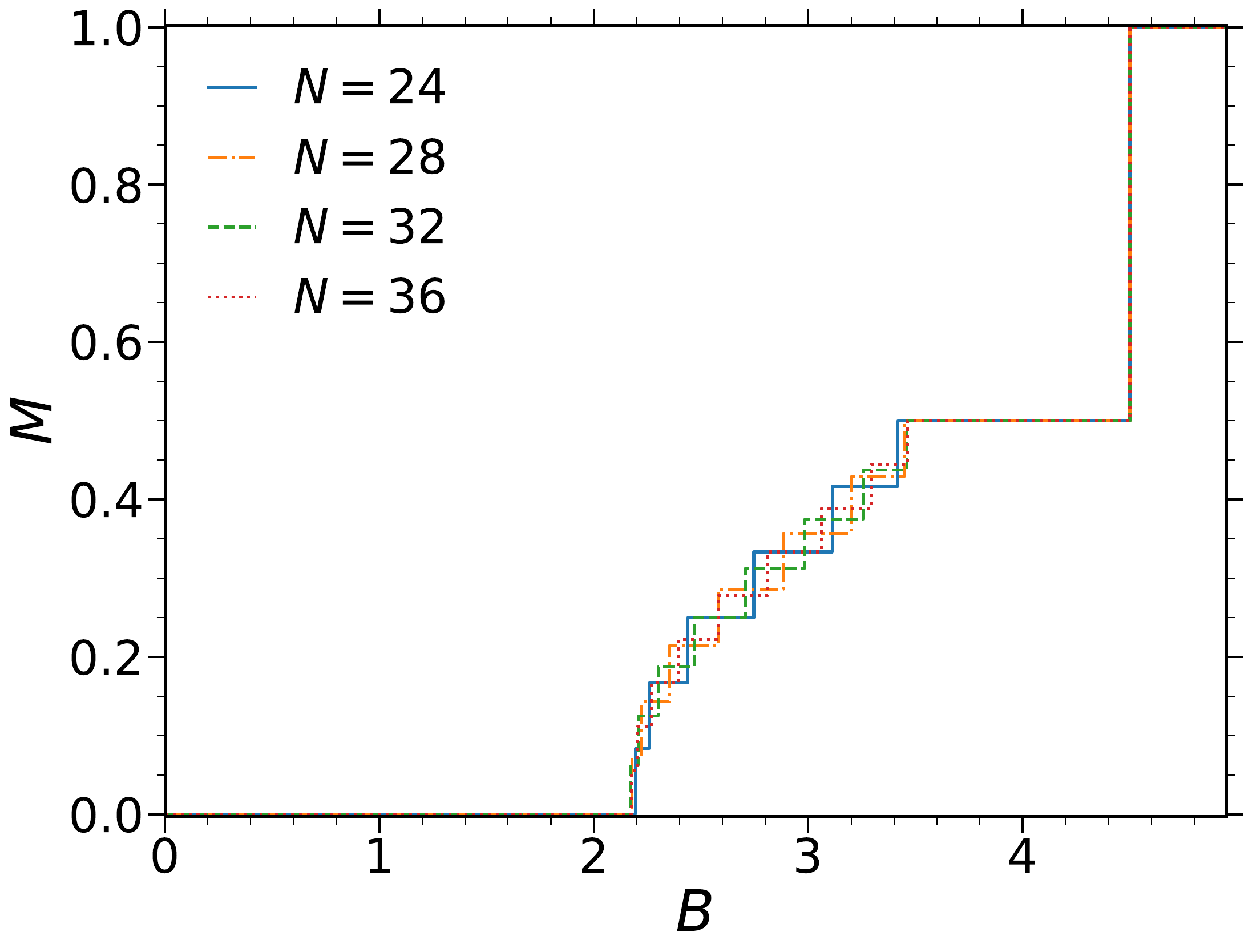}
      \caption{(Color online) Zero-temperature exact diagonalization
        results for the magnetization in the distorted sawtooth chain
        with KNB mechanism for the values of electric field magnitude,
        angle, $\theta_1$, $\theta_2$ given in the
        Eqs.~(\ref{eq:the_the_1})-(\ref{eq:the_the_3}). All three sets
        of parameters lead to the same magnetization curve as they all
        correspond to the same Hamiltonian (\ref{eq:ham_gen}), namely
        $J_1=1, J_2=1, J_3=3$, $\Delta_1=\Delta_2=\Delta_3=1$ and
        $D_1=1, D-2=-2, D_3=-1$.}
      \label{fig2}
    \end{center}
  \end{figure}

  Let us first check the cases of special directions of the electric
  field, which were considered previously for symmetric sawtooth chain
  in Ref.~\cite{sc_KNB}. There a remarkable result was obtained for
  the field parallel to the basal line (see Eq.~(\ref{eq:E_FB})), in
  which case no further constraints for the coupling constants are
  needed. Although one might expect similar features in the distorted
  chain as well, the situation turns out to be different. Namely for
  this specific case of an electric field $\bm{E}=(E, 0,0)$, the
  flat-band conditions immediately lead to the following DM
  parameters,
  \begin{eqnarray}
    D_1&=&0, \\
    D_2&=&-E\sin\theta_1, \nonumber \\
    D_3&=&E\sin\theta_2. \nonumber
  \end{eqnarray}
  Then a flat band results if the electric field has the magnitude and
  a symmetric exchanges have an appropriate ratio,
  \begin{eqnarray}\label{eq:phi=0}
    E &=& \pm
          \frac{1}{\sin\theta_2}
          \bigg[
          J_1^2\left(\frac{\sin^2\theta_2}{\sin^2\theta_1}+2\text{sign}(J_1)\Delta_1\frac{\sin\theta_2}{\sin\theta_1}+1\right)\nonumber\\
      &&~~~~~~~~~~~~
         -J_2^2\frac{\sin^2\theta_2}{\sin^2\theta_1}
         \bigg]^{\frac12}\,,\\
    J_3 &=& \frac{\sin\theta_2}{\sin\theta_1}J_2. \nonumber
  \end{eqnarray}
  Thus, the remarkable result of Ref.~\cite{sc_KNB} can be reproduced
  here only when $\theta_2=\theta_1$ and $J_3=J_2$, i.e., in the case
  of a non-distorted and symmetric sawtooth chain. The fact that the
  this electric-field driven flat band requires $J_3=J_2$ was found
  also in Ref.~\cite{JJJ}. Hence in general an additional constraint
  is needed, which was absent in the result of Ref.~\cite{sc_KNB} for
  $\phi=0$ only because $J_3=J_2$ was assumed at the outset. For the
  corresponding saturation field given in Eq.~(\ref{eq:EB0}) we find
  \begin{eqnarray}\label{eq:phi=0_B0}
    B_0&=&\frac{J_1}{2}\left(2\Delta_1+\frac{\sin^2\theta_1+\sin^2\theta_2}{\sin\theta_1\sin\theta_2}\right)\\
       &&~~+\frac{J_2}{2}\left(\Delta_2 + \frac{\sin\theta_2}{\sin\theta_1}\Delta_3\right),\nonumber
  \end{eqnarray}
  which reduces to the known results for the non-distorted sawtooth chain if  $\theta_2=\theta_1$ and $J_3\Delta_3=J_2\Delta_2$ \cite{sc_KNB, JJJ}, namely
  \begin{eqnarray}\label{eq:phi=0_B0_lim}
    B_0^{(0)}=J_1\left(1+\Delta_1\right)+J_2\Delta_2.
  \end{eqnarray}
  The saturation magnetisation has a non-monotonous dependence on
  the distorted bond angle $\theta_2$. For positive values of $J_1$
  it has a maximum at $\theta_2=\frac{\pi}{2}$ and minima
  at
  \begin{eqnarray}\label{eq:theta_min}
    \theta_{2}=\left|\arcsin\left(\frac{\sqrt{J_1}\sin\theta_1}{\sqrt{J_1+J_2\Delta_3}}\right)\right|.
  \end{eqnarray}
  Interestingly there is no dependence on $\theta_1$ in the saturation field in this case,
  \begin{eqnarray}\label{eq:phi=0_B0_1}
    B_0&=&\frac{J_1}{2}\left(2\Delta_1\pm\frac{2J_1^2+J_2\Delta_3}{\sqrt{J_1\left(J_1+J_2\Delta_3\right)}}\right)\\
    &&~~+\frac{J_2}{2}\left(\Delta_2 \pm \frac{\sqrt{J_1}}{\sqrt{J_1+J_2\Delta_3}}\Delta_3\right).\nonumber
  \end{eqnarray}
  It is also interesting to consider the case of small distortions,
  $\theta_2=\theta_1+\delta$, where $|\delta|\ll\theta_1$. In this
  case the saturation
  field depends on  $\delta$ in leading order as (for $\Delta_3=\Delta_2$)
  \begin{eqnarray}\label{eq:line_delta}
    B_0\simeq B_0^{(0)}+\frac 12 J_2\Delta_2 \cot\theta_1\cdot\delta.
  \end{eqnarray}
  In Ref.~\cite{sc_KNB}, an electric field pointing parallel to the
  zigzag bond, $\phi=\theta$, for the non-distorted sawtooth chain.
  In contrast to the previous case $\phi=0$ an additional constraint
  similar to the distorted case appeared,
  \begin{eqnarray}\label{eq:J3alpha}
    E&=&\pm\frac{2\sqrt{\cos^2\theta-a^2}}{a^2\sin\theta}\left|J_1\right|, \\
    J_2&=&J_3=\frac{2 \cos\theta}{a}J_1. \nonumber
  \end{eqnarray}
  To generalize this result to the distorted sawtooth case we choose
  the direction of the electric field parallel to the left side of
  triangles, $\phi=\theta_1$, which leads to the following
  configuration of the DM parameters:
  \begin{eqnarray}\label{eq:DDD_2}
    D_1&=&aE\sin\theta_1, \\
    D_2&=&0, \nonumber \\
    D_3&=&E\sin\left(\theta_1+\theta_2\right). \nonumber
  \end{eqnarray}
  The flat-band value of the electric-field magnitude and the
  additional constraint for the exchange couplings are then
  \begin{eqnarray}\label{eq:E_FB_2}
    E&=&\pm\frac{1}{\sin\left(\theta_1+\theta_2\right)}
         \Bigg[J_2^2 \left(\frac{\sin^2\left(\theta_1+\theta_2\right)}{a^2\sin^2\theta_1}-1\right)\\
     &&~~~~-J_1^2 \frac{\sin^2\left(\theta_1+\theta_2\right)}{a^2\sin^2\theta_1}-2J_1\left|J_2\right|\Delta_1\frac{\sin\left(\theta_1+\theta_2\right)}{a\sin\theta_1}\Bigg]^{\frac12}\nonumber\\
    J_3&=&\frac{\sin\left(\theta_1+\theta_2\right)}{a\sin\theta_1}J_1.\nonumber
  \end{eqnarray}
  The saturation field then takes the following form,
  \begin{eqnarray}\label{eq:B_0_2}
    B_0&=&\frac{J_2}{2}\left(\Delta_2+\frac{\sin\left(\theta_1+\theta_2\right)}{a\sin\theta_1}\right)\\
    &&~~~~+\frac{J_1\Delta_3}{2}\frac{\sin\left(\theta_1+\theta_2\right)}{a\sin\theta_1}.\nonumber
  \end{eqnarray}
  It is easy to see that this expression has no local minimum, and it
  has only a local maximum at
  $\theta_2=\theta_1+\frac{\pi}{2}$. However, the minimal value is
  reached at $\theta_2=\frac{\pi}{2}$. We conclude that for given
  value of $\theta_1$ the maximal possible value of the second angle
  is $\frac{\pi}{2}$. For $\theta_2=\frac{\pi}{2}$ the saturation field
  transforms to
  \begin{eqnarray}
    B_0=\frac{J_2}{2}\left(\Delta_2+\frac{\cot\theta_1}{a}\right)+\frac{J_1\Delta_3}{2}\frac{\cot\theta_1}{a}\,.
  \end{eqnarray}
  In the limit of small distortion this becomes
  \begin{eqnarray}\label{eq:line_delta_2}
    B_0\simeq B_0^{(0)}+\left(J_2+J_1\Delta_3\right)\frac{\cos\left(2\theta_1\right)}{\sin\theta_1}\cdot\delta\,,
  \end{eqnarray}
  where
  \begin{eqnarray}\label{eq:B_0_02}
    B_0^{(0)}=\frac{J_2}{2}\left(\Delta_2+\frac{2\cos\theta_1}{a}\right)+\frac{J_1\Delta_3}{2}\frac{2\cos\theta_1}{a}
  \end{eqnarray}
  is the saturation field for the non-distorted case \cite{sc_KNB,
    JJJ}. The situation when the electric field is parallel to the
  right sides of the triangle ($\phi=-\theta_2$ or
  $\phi=\pi-\theta_2$) is very similar, the flat-band value of
  electric field can be obtained form
  Eqs.~(\ref{eq:E_FB_2})-(\ref{eq:B_0_02}) by switching $J_2$ and
  $J_3$ and $\theta_1$ and $\theta_2$. To conclude, the result
  obtained in Ref.~\cite{sc_KNB} concerning the electric-field driven
  flat band without an additional condition for the exchange coupling
  constant was accidental in the sense that the couplings where chosen
  symmetric throughout, whereas this follows in the more general case
  from Eq.~(\ref{eq:phi=0}. However, as shown in Ref.~\cite{JJJ}, one
  can nevertheless pick a direction of the electric field which is not
  necessarily parallel to the lattice bonds and then adjust the
  magnitude of electric field to drive the system into the flat-band
  regime. For this one starts from the flat band constraints in terms
  of $D_1, D_2$, $D_3$ in Eqs.~(\ref{sol_y=ax})-(\ref{al_range}) and
  constructs a one-parameter family of mappings
  \begin{eqnarray}
    f_a: \begin{pmatrix}
      D_1 \\ D_2 \\ D_3
    \end{pmatrix}\rightarrow \begin{pmatrix}\phi \\ E \\ \theta\end{pmatrix}\,,
  \end{eqnarray}
  where the ratio $a$ of KNB constants plays the role of a free
  parameter. Thus, for given DM constants which correspond to a flat
  band in the general sawtooth chain, flat band induced by the KNB
  mechanism can be obtained from the electric field angle and
  magnitude and the bond angle as functions of $(D_1, D_2, D_3)$ and
  of the parameter $a$. These functions are given in
  Eq.~(\ref{eq:phi_E_2}). For the distorted sawtooth chain the mapping
  is more complicated and the second, distorted angle $\theta_2$ plays
  the role of an additional parameter, giving rise to a broader family
  of possible KNB realizations of the given solution of the flat band
  constraints. However, due to the symmetry of the KNB-induced DM
  parameters, there is an invariance under
  $\theta_1\leftrightarrow\theta_2$ and $D_2\leftrightarrow D_3$,
  i.e., the mapping where $\theta_1$ is a parameter instead of
  $\theta_2$ has a similar structure with $D_2$ and $D_3$
  interchanged,
  \begin{eqnarray}
    g_{a,\theta_2}: \begin{pmatrix}
      D_1 \\ D_2 \\ D_3
    \end{pmatrix}&\rightarrow& \begin{pmatrix}\phi \\ E \\ \theta_1\end{pmatrix},\\
    g_{a,\theta_1}: \begin{pmatrix}
      D_1 \\ D_2 \\ D_3
    \end{pmatrix}&\rightarrow& \begin{pmatrix}\phi \\ E \\ \theta_2\end{pmatrix},\nonumber
  \end{eqnarray}
  where
  \begin{eqnarray}
    g_{a,\theta_1}\circ P_{23}=g_{a,\theta_2}, \;\; P_{23}:\begin{pmatrix}
      D_1 \\ D_2 \\ D_3
    \end{pmatrix}= \begin{pmatrix}D_1 \\ D_3 \\ D_2\end{pmatrix}.
  \end{eqnarray}
  Let us illustrate this mapping for the example of the second case,
  when the left angle $\theta_2$ is expressed by the DM parameters and
  by the right angle $\theta_1$, i.e., $\theta_2$ $=$
  $\theta_2(D_1, D_2, D_3, a, \theta_1)$, so that the angle $\theta_1$
  plays the role of a free parameter here. From Eq.~(\ref{eq:DDD_def})
  we find
  \begin{eqnarray}\label{eq:DDD_th_th}
    D_2\sin\theta_2+D_3\sin\theta_1=\frac{D_1}{a}\sin\left(\theta_1+\theta_2\right)
  \end{eqnarray}
  which can be solved with respect to $\sin\theta_2$,
  \begin{eqnarray}\label{eq:theta_2}
    &&\theta_2(D_1, D_2, D_3, a, \theta_1)\\
    &&~~=\,\arcsin\frac{\text{sign}\left(a\right)D_1\sin\theta_1}{\sqrt{D_1^2+a^2 D_2^2-2 a D_1 D_2\cos\theta_1}}\nonumber
    \\
    &&~~~~~~-\arcsin\frac{\left|a\right|D_3\sin\theta_1}{\sqrt{D_1^2+a^2 D_2^2-2 a D_1 D_2\cos\theta_1}}.\nonumber
  \end{eqnarray}
  Combining the expressions for $D_2$ and $D_3$ from
  Eqs.~(\ref{eq:DDD_def}) and (\ref{eq:DDD_th_th}), one can then
  express angle and magnitude of the electric field
  \begin{eqnarray}\label{eq:Ephi}
    \phi&=&\arctan\frac{D_3\sin\theta_1+D_2\sin\theta_2(D_1, D_2, D_3, a, \theta_1)}{D_3\cos\theta_1-D_2\cos\theta_2(D_1, D_2, D_3, a, \theta_1)},\nonumber\\
    E&=&\pm\frac{\left|D_1\right|}{\left|a\right|}\bigg[1+\\
        &&~~~\left(\frac{D_3\cos\theta_1-D_2\cos\theta_2\left(D_1, D_2, D_3, a, \theta_1\right)}{D_3\sin\theta_1+D_2\sin\theta_2\left(D_1, D_2, D_3, a, \theta_1\right)}\right)^2\bigg]. \nonumber
  \end{eqnarray}
  Thus one of the bond angles, $\theta_1$ in this case, plays the role
  of a continuous parameter which leads to a flat band for given
  KNB-induced $D_1, D_2$ and $D_3$. Varying $\theta_1$ in
  Eq.~(\ref{eq:Ephi}) not keeps the flat band, but does not even
  change the values of the DM parameters. Let us exhibit two examples
  of this mapping, obtained in Ref.~\cite{JJJ}, both with
  $J_1=1, J_2=1, J_3=3$, $\Delta_1=\Delta_2=\Delta_3=1$,
  $D_1=1, D_2=-2, D_3=-1$, and two distinct values of the right bond
  angle, $\theta_1=\frac{\pi}{8}$ and $\theta_1=\frac{\pi}{2}$, whivh
  both correspond to the same Hamiltonian (\ref{eq:ham_gen}). The
  parameters of the electric field and the left bond angle are
  nevertheless different,
  \begin{eqnarray}\label{eq:the_the_1}
    \theta_1&=&\frac{\pi}{8}, \;\; \theta_2=2\arcsin\left[\frac{\sin\frac{\pi}{8}}{\sqrt{5-4\cos\frac{\pi}{8}}}\right],\\
    \phi&=&\arctan\frac{\left(5-4\cos\frac{\pi}{8}\right)\left(\sin\frac{\pi}{8}-2\sin\theta_2\right)}
            {3\left(2-\cos\frac{\pi}{8}\right)},\nonumber\\
    E&=&\pm\frac{\sqrt{5+4\cos\left(\frac{\pi}{8}+\theta_2\right)}}{\left|\sin\frac{\pi}{8}-2\sin\theta_2\right|},\nonumber
  \end{eqnarray}
  and
  \begin{eqnarray}\label{eq:the_the_2}
    \theta_1&=&\frac{\pi}{4},\;\; \theta_2=2\arcsin\left(\frac{1}{\sqrt{10-4\sqrt 2}}\right),\\
    \phi&=&\arctan\frac{17\left(\sqrt 2-4\sin\theta_2\right)}{48+9\sqrt 2},\nonumber\\
    E&=&\pm\sqrt{1+\left(\frac{1+2\sqrt 2\cos\theta_2}{1-2\sqrt 2\sin\theta_2}\right)^2} \nonumber
  \end{eqnarray}
  Interestingly, the solution with $\theta_1=\frac{\pi}{3}$ corresponds
  to a non-distorted sawtooth chain,
  \begin{eqnarray}\label{eq:the_the_3}
    \theta_1=\frac{\pi}{3},
    \;\;
    \theta_2=\frac{\pi}{3},
    \;\;
    \phi=-\frac{\pi}{6},
    \;\;
    E=\pm 2.
  \end{eqnarray}
  These three sets of the parameters for the sawtooth chain with KNB
  parameters all realize the same physical picture. To illustrate this
  we calculated the zero-temperature magnetization curves by exact
  diagonalization (see Fig.~\ref{fig2}). For a magnetic field below
  saturation, the localized magnons of the flat bands lower the energy
  of the system according to Eq.~(\ref{eq:one-m_spec}). Thus, a magnon
  crystal forms instantly during the magnetization process giving rise
  to a magnetization plateau at half the saturated magnetization
  equal. However, at $B=B_0$ this magnon crystal becomes degenerate
  with the saturated fully polarized spin configuration. This leads to
  a magnetization jump from the $M=\frac 12$ to the $M=1$
  plateau. This magnetization profile is typical for systems with
  localized magnons \cite{der15,schul02,rich04,rich05} and also
  observed in the Fig.~\ref{fig2}. An additional plateau at $M=0$ is
  in fact independent of the flat band and rather indicates the
  spin-liquid ground state which is inherent to the sawtooth chain
  with DM interactions \cite{mila}.

  \section{Summary}

  In the present paper we presented further analysis of the solutions
  of flat-band constraints for the a generalized sawtooth chain model
  with DM interactions. Our main focus was the mapping of the general
  solution in terms of DM constants to the case when the effective DM
  interaction is generated by the KNB mechanism of spin-induced
  ferroelectricity, which is sensitive to the geometry. In contrast to
  our previous results given in Refs.~\cite{sc_KNB} and \cite{JJJ},
  here we considered distorted sawtooth chain featuring two distinct
  bond angles at the base of the triangles. The distortion is related
  to the modification of relations linking DM parameters with the
  electric field characteristics, magnitude and direction. We
  considered in particular a electric field parallel to the lattice
  bonds. For each cases we found the corresponding values of the
  electric field driving the one-magnon spectrum into a flat band
  mode. The behavior of the saturation magnetic field, particularly
  its dependence on the distorted angle, was also analyzed. As a main
  result we constructed the mapping from the general solution of the
  flat band constraints, given in terms of general DM parameters, into
  the electric-field driven flat-band constraints, by expressing
  electric field magnitude, angle and one of the bond angles in terms
  of $D_1, D_2$, and $D_3$.

  \section*{Acknowledgements}

  V.O. is grateful to the Center for Electronic Correlations and
  Magnetism for the hospitality during his visits to the University of
  Augsburg; he and L.A. also acknowledge partial financial support
  from ANSEF (Grants No. PS-condmatth-3273) and CS RA MESCS (Grants
  No. 21AG-1C047).  V.O. and M.K. acknowledge partial funding by
  Deutsche Forschungsgemeinschaft (DFG, German Research Foundation) --
  TRR360 -- 492547816.

\end{document}